\documentclass{emulateapj}
\usepackage{epsfig}

\newcommand{\saxj}{\mbox{SAX J1808.4$-$3658}}
\newcommand{\RXTE}{\textit{RXTE}}
\newcommand{\us}{$\mu$s}

\newcommand{\fluxunits}{~erg~cm$^{-2}$~s$^{-1}$}

\newcommand{\SpinPaper}{H08}

\begin{document}

\title{The Luminosity and Energy Dependence of Pulse Phase Lags in the
Accretion-powered Millisecond Pulsar SAX~J1808.4$-$3658}
\shorttitle{Soft lags in \saxj}
\shortauthors{Hartman et~al.}

\submitted{Accepted by ApJ}

\author{Jacob M. Hartman\altaffilmark{1,2,3},
Anna L. Watts\altaffilmark{4}, and Deepto Chakrabarty\altaffilmark{1}.}
\altaffiltext{1}{Kavli Institute for Astrophysics and Space Research,
Massachusetts Institute of Technology, Cambridge, MA 02139}
\altaffiltext{2}{National Research Council research associate}
\altaffiltext{3}{New address: Code 7655, Naval Research Laboratory,
Washington, DC 20375; Jacob.Hartman@nrl.navy.mil}
\altaffiltext{4}{Astronomical Institute ``Anton Pannekoek'',
University of Amsterdam, Kruislaan 403, 1098 SJ Amsterdam, Netherlands}

\begin{abstract}

Soft phase lags, in which X-ray pulses in lower energy bands arrive later than
pulses in higher energy bands, have been observed in nearly all
accretion-powered millisecond pulsars, but their origin remains an open
question.  In a study of the 2.5~ms accretion-powered pulsar \saxj, we report
that the magnitude of these lags is strongly dependent on the accretion rate.
During the brightest stage of the outbursts from this source, the lags
increase in magnitude as the accretion rate drops; when the outbursts enter
their dimmer flaring-tail stage, the relationship reverses.  We evaluate this
complex dependence in the context of two theoretical models for the lags, one
relying on the scattering of photons by the accretion disk and the other
invoking a two-component model for the photon emission.  In both cases, the
turnover suggests that we are observing the source transitioning into the
``propeller'' accretion regime.

\end{abstract}

\section{Introduction}

Accretion-powered millisecond pulsars (AMPs) provide a unique laboratory for
studying the process of disk accretion onto magnetic stars
(\citealt{Psaltis99} and refrences therein).  The $\sim$$10^8$~G stellar
magnetic field truncates the accretion disk only a few neutron star radii
above the surface, channeling the infalling matter toward the magnetic poles.
The resulting X-ray--emitting hot spots and accretion shocks are modulated by
the stellar rotation to produce the observed pulsations.  Such compact
accretion geometry poses a particularly interesting challenge.  We want to
understand the details of the interaction between the magnetic field and the
disk, and the effects that result from the proximity of different emitting
regions.  The accretion-powered 2.5~ms pulsar \saxj\ is a particularly good
system for this type of study: in addition to having multiple outbursts, it
spans orders of magnitude in accretion rate.  In this paper we report our
study of the energy-dependence of pulsations from this source and show that it
gives new insight into these problems.

\saxj\ was the first detected AMP \citep{Wijnands98}, and it remains the best
studied.  Persistent pulsations have been detected throughout the four $\sim$1
month outbursts from this source that have been observed by the {\em Rossi
X-ray Timing Explorer} (\RXTE) during 1998--2005.  The timing of these
pulsations establishes a 2.01~hr binary orbital period \citep{Chakrabarty98}
that is increasing on a timescale of $P_{\rm{orb}}/\dot{P}_{\rm{orb}} =
(66\pm4)\times10^6$~yr, possibly due to the ablation of its low-mass companion
during X-ray quiescence (\citealt{Hartman08}, hereafter \SpinPaper;
\citealt{DiSalvo08}).  The presence of X-ray pulsations during the peak
accretion rate and the long-term spin-down of this source provide a tight
constraint on its magnetic field: $B = (0.4$--$1.5)\times10^8$~G (\SpinPaper;
\citealt{Psaltis99}).  A distance of 3.4--3.6~kpc is estimated from
thermonuclear X-ray burst properties \citep{Galloway06}.

The outbursts from \saxj\ had similar light curves, which we divide for
convenience into four stages (see Fig.~2 of \SpinPaper\ for a schematic): a
rise from quiescence lasting $\approx$5~days; a short-lived peak at a
2--25~keV flux of $(1.9$--$2.6)\times10^{-9}$\fluxunits; a 10--15~day slow
decay in luminosity that lasts until the source reaches
$\approx$$0.7\times10^{-9}$\fluxunits; and finally a rapid fall in luminosity
followed by weeks of low-luminosity flares with a $\sim$5~d periodicity,
during which the observed flux can vary by a factor of 1000
\citep{Wijnands01}.  This sudden drop has been interpreted as a transition
into a ``propeller'' accretion regime \citep[e.g.,][]{Gilfanov98} in which
accretion is only partially inhibited \citep{Spruit93, Rappaport04}, although
it may also be caused by a change in the viscosity of the accretion disk as it
transitions into quiescence \citep{Gilfanov98, Gierlinski02}.

Shortly after the discovery of pulsations in \saxj, \citet{Cui98} noted the
presence of soft phase lags, the tendency of the pulses at lower (``softer'')
energies to arrive later than the corresponding pulses at higher (``harder'')
energies.  The magnitude of the lags increased approximately linearly between
2--10~keV and saturated at 0.08 rotational cycles (200~\us) above 10~keV.
Similar soft lags have been detected in six other AMPs:  IGR~J00291$+$5934
\citep{Galloway05}, XTE~J0929$-$314 \citep{Galloway02}, XTE~J1751$-$305
\citep{Gierlinski05}, XTE~J1807$-$294 \citep{Kirsch04}, XTE~J1814$-$338
\citep{Watts06}, and HETE~J1900.1$-$2455 \citep{Galloway07}.

A number of explanations for the AMP soft lags have been proposed.  Models
that include only Doppler boosting, due to the stellar rotation
\citep{Ford00,Weinberg01} or the orbiting material in the accretion disk
\citep{Sazonov01}, have shortcomings that are summarized by
\citet{Gierlinski02}.  The two most plausible explanations invoke a
two-component emission model with differing angular dependence and spectra
(\citealt{Poutanen03}; hereafter PG03) or the scattering delay of hot photons
from the accretion shock off the cooler disk or stellar surface
(\citealt{Falanga07}; hereafter FT07).  In this paper we present an
energy-resolved analysis of the persistent pulsations for all the outbursts of
\saxj\ detected by \RXTE.  We will discuss the implications of our results for
these models of the soft lags.

\section{Data analysis}

For our study of the energy dependence of the persistent pulsations, we
analyzed the same \RXTE\ observations of \saxj\ as in \SpinPaper\ (refer to
Table~1 therein).  The \verb!E_125US_64M_0_1S! data mode of the \RXTE\
Proportional Counter Array (PCA; \citealt{Jahoda96}) was used for nearly all
the observations.  It provides 122~\us\ time bins and 64~energy channels
divided evenly over the detector response, a good combination for discerning
the energy dependence of the persistent pulsations.  The energy band covered
by each channel varied over the 7~yr of observation spanned by this study due
to adjustments to the PCA anode voltages \citep{Jahoda06}; we accounted for
this effect throughout our measurements.

For our timing analysis, we shifted the photon arrival times to the solar
system barycenter, applied the \RXTE\ fine clock correction, and filtered out
data during Earth occultations and intervals of unstable pointing.  We also
removed any data within 5~min of thermonuclear X-ray bursts.  To measure the
phases and fractional amplitudes of the persistent pulsations for a given
energy band $E$, we folded 512~s intervals of data using the phase timing
solution derived in \SpinPaper.  Dividing the resulting pulse profiles into
$n$ phase bins with photon counts $x_{E,j}$, $j = 1 \dots n$, the amplitude
($A_{E,k}$) and pulse phase offset ($\phi_{E,k}$) for the $k$th
harmonic\footnote{Throughout this paper, we number the harmonics such that the
$k$th harmonic is $k$ times the frequency of the 401~Hz fundamental.} is
\begin{equation}
  A_{E,k} e^{2\pi ik\phi_{E,k}} = 2 \sum_{j=1}^{n} x_{E,j} e^{2\pi ijk/n} \, .
  \label{eq:Folding}
\end{equation}
We measure the phase offsets of the energy bands relative to the softest band,
$E_0$: $\Delta\phi_{E,k} = \phi_{E,k} - \phi_{E_0,k}$.  Thus negative values
of $\Delta\phi_{E,k}$ indicate that pulsations of the $k$th harmonic within
band $E$ arrived earlier than the pulsations within the softest band.  The
uncertainties of these phase offsets due to Poisson noise are
\begin{equation}
  \sigma_{{\rm E},k} = \frac{\sqrt{2 N_E}}{2\pi k A_{E,k}} \, ,
\end{equation}
where $N_E$ is the total number of photons in the energy band (\SpinPaper).
For our analysis, we rejected phase measurements with $\sigma_{{\rm E},k} >
0.04$~cycles (i.e., 0.1~ms).

The measured fractional rms amplitudes are
\begin{equation}
  r_{E,k} = \frac{A_{E,k}}{\sqrt{2}\left(N_E - B_E\right)} \, ,
\end{equation}
where $B_E$ is the approximate number of background events within our energy
band and time interval, estimated using the FTOOL {\tt pcabackest}.\footnote{
\url{http://heasarc.gsfc.nasa.gov/docs/xte/recipes/pcabackest.html}}
Uncertainties on the fractional amplitudes are computed using the method
described by \citet{Groth75} and \cite{Vaughan94}, which correctly accounts
for the addition of noise to the complex amplitude of the signal.

We estimated the 2.5--25~keV fluxes of the persistent emission by fitting the
spectra extracted from Standard2 mode data (128 energy channels, 16~s time
resolution) with an absorbed blackbody plus power law model, including an iron
line if there were significant residuals around 6.4~keV; see \citet{BurstDB},
\S2.1, for full details.

\section{Results}

\begin{figure}[t!]
  \begin{center}
    \includegraphics[width=0.45\textwidth]{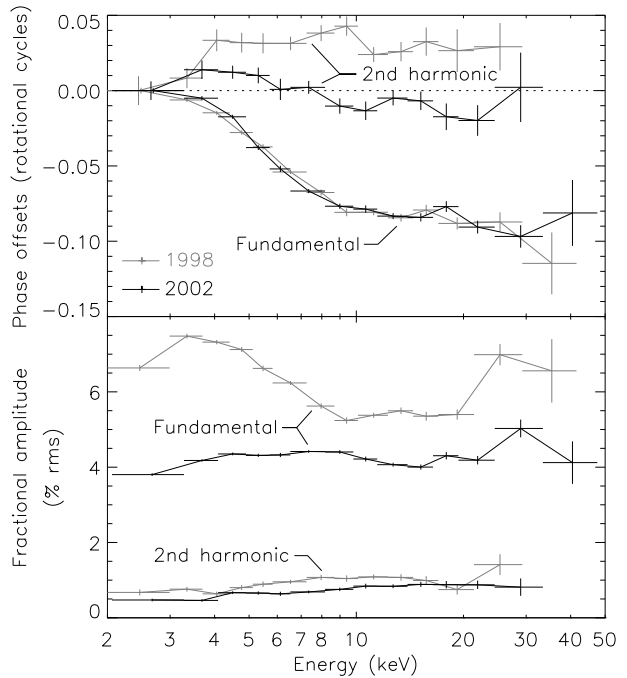}
  \end{center}
  \caption{ Average phase offsets $\Delta\phi_{E,k}$ (relative to the 2.1--3.3
    keV band) and fractional amplitudes during the slow-decay stages of the
    2002 outburst (black; MJD 52564.1--52575.0) and the 1998 outburst (grey;
    MJD 50919.0--50932.8, the same pointed observations used by
    \citealt{Cui98}).  Negative offsets indicate soft lags.
  \label{fig:AvgPulsationsVsEnergy}}
\end{figure}

\begin{figure*}[t!]
  \begin{center}
    \includegraphics[width=1\textwidth]{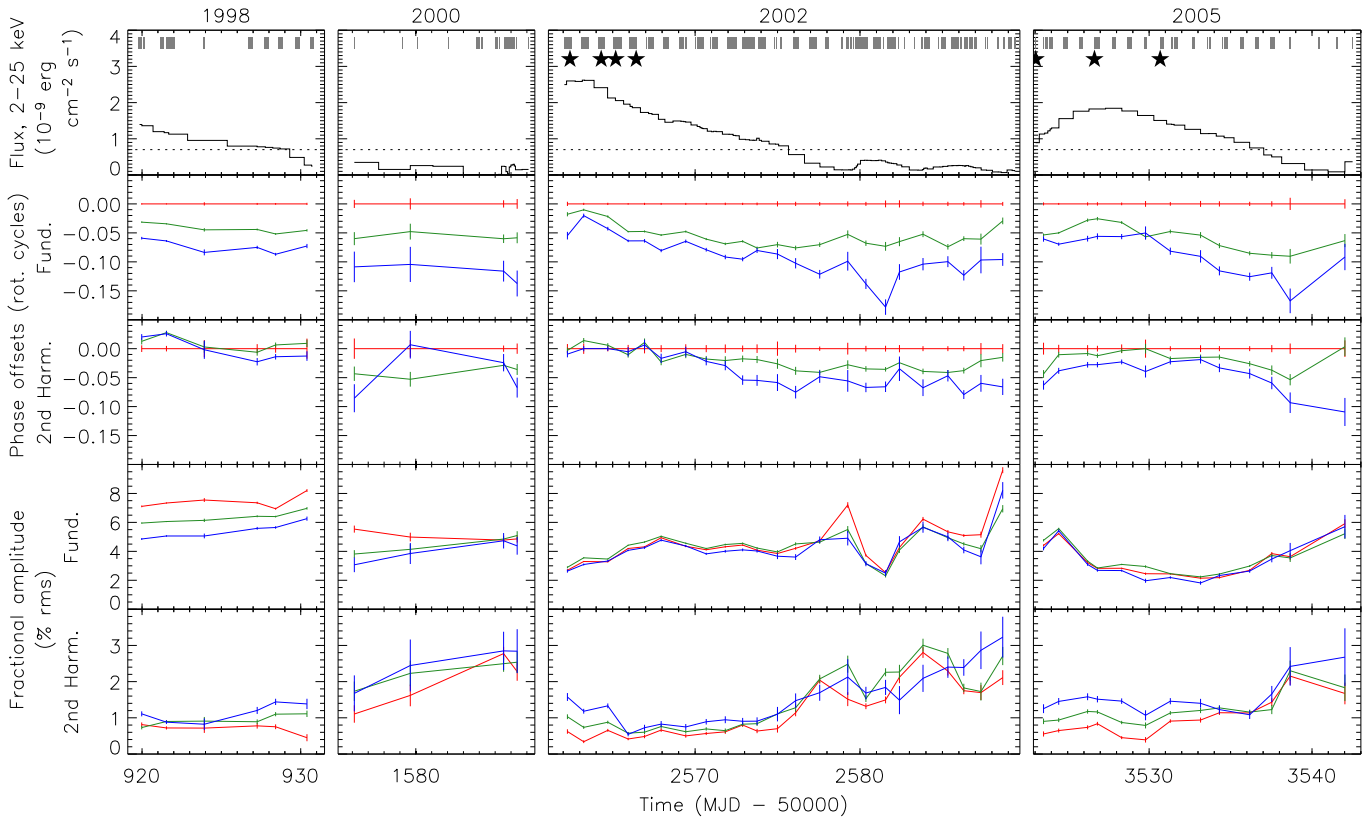}
  \end{center}
  \caption{ Phase offsets and fractional amplitudes as a function of energy
    for the four \saxj\ outbursts observed by \RXTE.  The top panels show the
    background-subtracted light curves for each outburst.  The strips along
    the tops of the graphs indicate the times of observations; stars indicate
    the times of the seven observed thermonuclear bursts.  The dotted line
    indicates the critical flux of $0.7\times10^{-9}$\fluxunits, at which the
    source transitions into the flaring-tail stage of the outburst.  The
    second and third panels from the top show the phase offsets of the medium
    energy band (green; roughly 5--10~keV) and hard energy band (blue;
    $\approx$10--20~keV) pulsations relative to the soft band (red;
    $\approx$2--5~keV).  Negative phase offsets indicate that higher-energy
    pulses lead the soft pulses.  The bottom two panels show the
    background-subtracted fractional amplitudes for these three energy bands.
    Each point in the phase and fractional amplitude plots gives a day-long
    average.  (They are not evenly spaced at 1~d intervals due to the uneven
    spacing of the \RXTE\ observations.)  Data within 5~min of thermonuclear
    bursts were excluded from these averages.
  \label{fig:BigEnergyPlot}}
\end{figure*}

To assess the energy dependence of the pulse phase and fractional amplitudes,
we calculated the average phase offsets and fractional amplitudes during the
11~d slow-decay stage of the 2002 outburst.
Figure~\ref{fig:AvgPulsationsVsEnergy} shows the results, along with
measurements from the 1998 outburst for comparison.  We chose this time range
because it was the longest and most data-rich interval during which the pulse
profile of \saxj\ was observed to be reasonably stable.  The pulse profiles
during the slow-decay stage of the 2005 outburst, although similar to the 2002
profiles, were somewhat less stable (see Fig.~3 of \SpinPaper).

The soft phase lag of the pulsations is immediately apparent for the
fundamental.  Between 2~keV and 10~keV, the pulses in harder bands arrive
progressively sooner, with a slope of $-0.014\pm0.002$~cycles~keV$^{-1}$.
Between 10~keV and 20~keV, this trend saturates at a lead of 0.08~cycles
(200~\us).  There is impressive agreement between the 1998 and 2002 outbursts,
which also had nearly identical light curves; the phase lags during the dimmer
2005 outburst are similar in shape but plateau at a smaller magnitude.  The
phase offsets of the second harmonic are less pronounced, disappearing almost
entirely or becoming hard lags when the source is bright, as is the case for
the data in Figure~\ref{fig:AvgPulsationsVsEnergy}.

In contrast, the energy dependence of the fractional amplitudes varies
considerably between outbursts.  The fractional amplitude of the fundamental
during the slow-decay stage of the 1998 outburst (5.2--7.5\% rms) was
considerably higher than during the 2002 outburst (3.8--5.0\% rms), as can be
seen in Figure~\ref{fig:AvgPulsationsVsEnergy}; the amplitude of the 2005
outburst (2.0--2.6\% rms) was lower still.  The morphology of the energy
dependence differs as well.  The 1998 outburst exhibits a large fractional
amplitude peak at 3.5~keV and falling amplitudes up to 20~keV, similar to what
is seen in XTE~J1814$-$338 \citep{Watts06}.  In contrast, during 2002 and 2005
the fractional amplitude vs. energy curves are much flatter, with small but
significant peaks at 8~keV.  Given the strong similarities between the light
curves of the 1998 and 2002 outbursts, it is curious that the energy
dependence of their fractional amplitudes is markedly different.  All three
outbursts show weak evidence of the fundamental's fractional amplitude
increasing again above $\approx$20~keV.  The fractional amplitude of the
second harmonic generally increases with energy for all the outbursts,
indicating less sinusoidal pulse profiles in harder energy bands.  XMM
observations of the 2008 outburst of this source reveal that the fractional
amplitude of both harmonics continues to decrease below 2~keV
\citep{Patruno09b}.

Figure~\ref{fig:BigEnergyPlot} illustrates how the energy-dependent phases and
fractional amplitudes change over the course of the four observed outbursts.
The soft lags in the fundamental are apparent throughout all four outbursts;
for the second harmonic, soft lags are apparent everywhere except during the
outburst peaks.  However, the degree of these lags is not constant: they are
much more pronounced during the tails of the outbursts, when the flux is low.
\citet{Watts06} noted a similar pattern in XTE~J1814$-$338, although that
source did not trace as wide a range of fluxes, so the relationship was less
clear.  The energy dependence of the fractional amplitudes does not exhibit
any clear correlation with the source flux or any other measured parameters,
apart from the general trend of fractional amplitudes increasing in the
outburst tails (see \S3.4 of \SpinPaper).

\begin{figure}[t!]
  \begin{center}
    \includegraphics[width=0.40\textwidth]{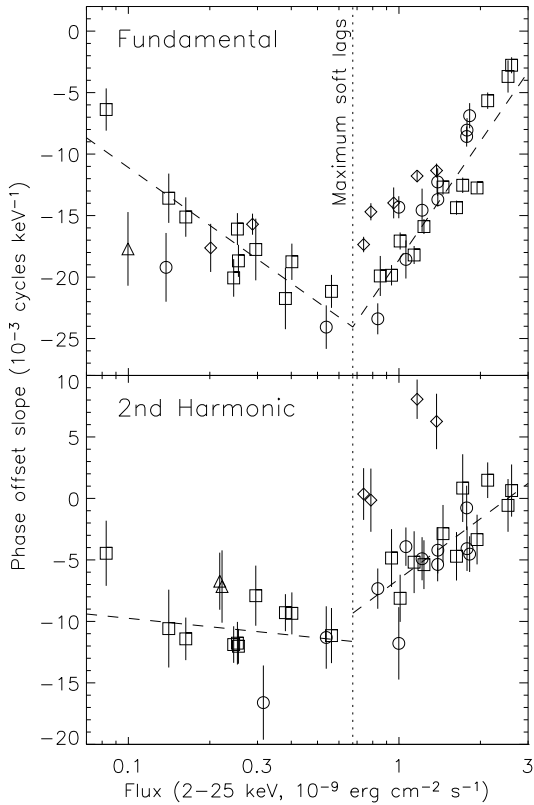}
  \end{center}
  \caption{ Correlation between the slope of the phase lags and the flux in
    \saxj.  Each slope is determined by a linear fit to the phase offset
    vs. energy measurements over 2--10 keV, with each point representing a 1~d
    average.  Large negative slopes indicate large soft lags.  The data from
    all four outbursts are shown: 1998 as diamonds, 2000 as triangles, 2002 as
    squares, and 2005 as circles.  The dashed lines indicate the best linear
    fits above and below the $0.68\times10^{-9}$\fluxunits\ turnover.  These
    fits are based on the 2002 and 2005 data only, as the 1998 and 2000
    outbursts exhibit greater scatter.
  \label{fig:PhaseLagSlopeVsFlux}}
\end{figure}

To further understand the relationship between the soft lags and the flux, we
measured the phase vs. energy slopes over 2--10~keV.  We chose this range
because it has the most significant energy dependence for the fundamental and,
to a lesser extent, for the second harmonic.
Figure~\ref{fig:PhaseLagSlopeVsFlux} plots the resulting slopes as a function
of flux.  For the fundamental, the magnitude of the soft lags is clearly
anticorrelated with the 2--25 keV flux when the source is bright: as the
source dims from its peak at $2.7\times10^{-9}$\fluxunits\ down to roughly
$0.7\times10^{-9}$\fluxunits, the magnitude of the soft lags grows.  This
range of fluxes covers the peak and slow-decay stages of the outburst.  At
lower fluxes, the relation reverses, and the magnitude of the lags becomes
correlated with flux.  The flux at which this turnover occurs is significant:
once it falls to $0.7\times10^{-9}$\fluxunits, the source dims rapidly and
enters the flaring-tail stage of the outbursts.

The trend for the second harmonic is somewhat less clear.  Measuring a slope
through the 2--10~keV fractional amplitude points does not work as well since
the energy dependence is less linear (top panel of
Fig.~\ref{fig:AvgPulsationsVsEnergy}), so there is more scatter.  Even so, the
data strongly suggest a similar relationship at high fluxes, particularly when
the relatively scattered 1998 data are excluded.

The slopes of these trends differ significantly for the two harmonics.  In the
low-flux regime, the slopes are $-16\pm2$ for the fundamental and $-2\pm2$ for
the second harmonic; at high fluxes, they are $32\pm4$ for the fundamental and
$17\pm3$ for the harmonic.  (The units in all cases are
$10^{-3}$~cycles~keV$^{-1}$~[log~flux]$^{-1}$.)  These slopes were fit using
the 2002 and 2005 data only, as the two earlier outbursts had appreciably more
scatter and less data.  Note that these units reflect {\em rotational} cycles
of the star, so a mechanism that induces the same phase lag in all photons of
a given energy would result in the slopes being equal for the two harmonics.
For the high-flux slopes, we can conclude with 8$\;\sigma$ significance that
this is not the case.

We also looked for correlations between the lags and the spectral parameters
from our blackbody plus power law fits.  The only parameter for which a
correlation is apparent is the normalization of the hard power law component.
This relation is not surprising, because the hard normalization and the total
X-ray flux are linearly proportional: during the 1998--2005 outbursts,
$\log(N_{\rm hard}) = (1.02\pm0.03) \log(f_x)$, with a reduced $\chi^2$ near
unity.  The uncertainties of $N_{\rm hard}$ are much greater than the flux
uncertainties, particularly when the source is dim, and the large scatter
obscures whether a turnover is present in the lag vs. $N_{\rm hard}$ relation.

\section{Discussion}

The energy dependence of the pulsations from \saxj\ exhibit strong variability
both between and during the four outbursts observed by \RXTE.  The overall
fractional amplitudes and the shape of the amplitude vs. energy curves differ
substantially between outbursts, with no obvious correlation with flux; the
only clear trend is that the pulsations generally become less sinusoidal with
energy in the observed 2--30~keV range.  In contrast, the soft lags of the
pulses have a consistent energy dependence, increasing in magnitude in the
2--10~keV band and saturating at 10--30~keV, with a strong dependence on flux.
At the highest fluxes the lags are near zero, but they increase steadily as
the flux falls.  Below a 2--25~keV flux of $\approx$$0.7\times
10^{-9}$\fluxunits, this trend reverses and lags get smaller again.  This
particular flux also marks the point in the outburst where the luminosity
decays rapidly and the source enters the flaring tail.

\begin{figure}[t!]
  \begin{center}
    \includegraphics[width=0.47\textwidth]{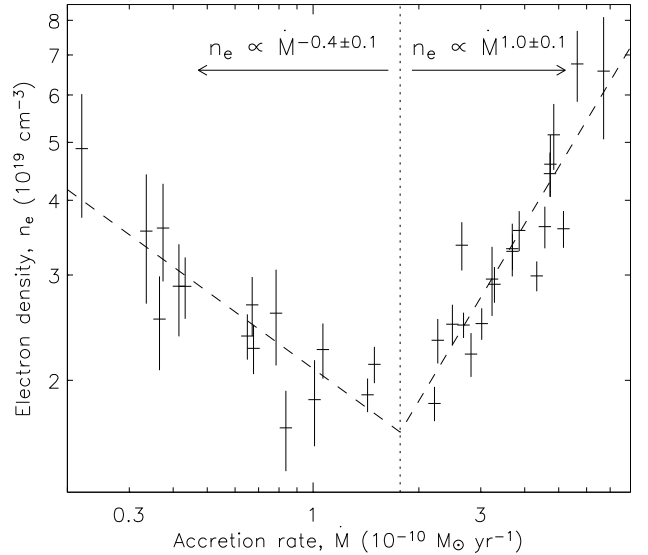}
  \end{center}
  \caption{ Electron density $n_e$ of the inner accretion disk, as fit using
    the model of FT07, versus the mass accretion rate for the 2002 and 2005
    outbursts of \saxj.  The vertical dotted line marks the $\dot{M}$ at which
    the magnitude of the soft lags is greatest and below which the outburst is
    in its flaring-tail stage; a transition from normal accretion to the
    ``propeller'' regime can explain this change in the observed behavior.
    The dashed lines give the best fits of $n_e \propto \dot{M}^\gamma$ in
    each region.
  \label{fig:FalangaFitVsMDot}}
\end{figure}

Soft lags are a common feature of AMP pulsations, and two plausible models
have emerged to explain their existence.  PG03 suggested a two-component model
involving soft emission from a surface hot spot and a hard Comptonized
component from the shocked layer in the accretion funnel.  Each component has
a different beaming pattern, which is affected in a different way by the
Doppler effect.  This results in the soft pulse lagging the hard pulse.  While
this model works well for \saxj\ and XTE~J1751$-$305 \citep{Gierlinski05}, it
does not predict the reduction in soft lags at very high energies seen in
IGR~J00291$+$5934 \citep{Falanga05}.  As a result FT07 proposed an alternative
model in which the disk plays a prominent role.  In this model, the main
pulsed element is the hard component from the accretion column: this radiation
is then downscattered by cold plasma in the inner disk (or on the neutron star
surface) to generate a soft lagging component.\footnote{\citet{Ibragimov09}
recently argued that photoelectric absorption in the accretion disk would
prevent the large number of scattering events required by the FT07 model.
However, standard disk models give a temperature of $\sim$300~eV in the inner
disk \citep{Shakura73}; the actual temperature will be higher due to the
irradiation.  At these temperatures and the expected electron densities ($n_e
\sim 3\times10^{19}$~cm$^{-3}$; see FT07 and Fig.~\ref{fig:FalangaFitVsMDot}
here), all elements through oxygen will easily be fully ionized by the hard
tail of the thermal distribution.  The only significant source of
photoelectric absorption is Fe, but for this $n_e$ only Fe~{\sc xxv} and
higher remain at $T \ge 300$~eV, with the transition to full ionization
occuring at 400--600~eV \citep{Rubiano04}.  Its He-like and H-like forms have
K edges at 9~keV, so they will not absorb the softer photons responsible for
the lags.  Since the FT07 model is not ruled out {\em a priori}, this
discussion attempts to address both models without preference.}  These soft
lags get larger as the density of the scattering plasma falls.

Let us now explore the factors that might cause the flux dependence of the
soft lags, assuming that flux directly traces accretion rate.  We then need to
ask how the properties of the system are expected to change as the accretion
rate falls.  For a star with a magnetic field that is strong enough to channel
the accretion flow, the reduction in accretion rate should cause an increase
in inner disk radius (e.g., \citealt{Ghosh77}).  This will have two relevant
effects.  First, the density of scattering plasma near the star will fall:
within the model of FT07 this would cause the soft lags to increase.  In
addition the relative angular velocity of the inner disk and star will reduce.
This will affect the angle at which the accretion funnel impacts the surface
\citep{Romanova04}, changing the column density above any surface hot spot.
This type of change might work within the model of PG03 to affect the soft
lags.

If these changes continue steadily, however, it is hard to see what could
cause the turnover seen in Figure~\ref{fig:PhaseLagSlopeVsFlux}.  One
intriguing possibility that would give a very natural explanation for such a
turnover is that the system reaches the point where the magnetospheric radius
$r_m$ exceeds the corotation radius $r_c$.  This is often referred to as the
``propeller'' regime, since it was originally thought that matter would then
be ejected from the system, with the star going from spin-up to spin-down
\citep{Illarionov75}.  In fact, the transition is probably not sharp, and
$r_m$ needs to be substantially larger than $r_c$ for ejection to occur
\citep{Spruit93}.  At intermediate accretion rates, continued accretion is
possible even while the neutron star starts to spin down \citep{Spruit93,
Rappaport04}.  Simulations by \citet{Long05} indicate that the transition
between spin up and spin down should occur when $r_m / r_c \approx 0.7$ for
AMPs.  The observed turnover happens at an accretion rate of $\dot{M} =
1.8\times10^{-10}\ M_\sun$~yr$^{-1}$, giving $r_m / r_c = 0.6$ for the
magnetic field $B = 1.5\times10^8$~G from \SpinPaper.\footnote{Accretion rates
were estimated in the usual manner, by equating the bolometric and infall
luminosities: $4\pi d^2 f_x c_{\rm{bol}} = GM\dot{M}/R$, using $d = 3.5$~kpc
and the bolometric correction $c_{\rm{bol}} = 2.12$ from \citet{Galloway06}.
The Alfv\'en radius was estimated as $r_m = k_A (2GM)^{-1/7} \dot{M}^{-2/7}
(BR^3)^{4/7}$, with $k_A = 0.5$ as per \citet{Long05}.  $M = 1.4\ M_\sun$ and
$R = 10$~km were used in all cases.}  This close agreement between the
simulations and the $\dot{M}$ at which the flux rapidly drops and the soft
lags turnover strongly suggests that \saxj\ passes through this transition as
the accretion rate falls.  In this eventuality there are two effects that
could cause the turnover in soft lags.

The first is due to the fact that the relative angular velocity of the inner
disk and the star changes sign.  The accretion funnel will go from being
dragged forwards to being dragged backwards (assuming that it settles rapidly;
\citealt{Romanova04}).  The column depth above any surface hot spot should
therefore go through a maximum near the zero torque point, causing a turnover
in the scattering properties of the accretion column.  If the accretion column
plays a major role as discussed in PG03, this could cause the observed
turnover.  Another explanation within the PG03 model is the suppression of the
lagged soft component at high fluxes due to the increased optical depth of the
accretion shock when $\dot{M}$ is large.  However, this mechanism would
require the normalization of the soft component to decrease during the
outburst peaks, contrary to what is seen (cf. Fig.~2 of
\citealt{Gierlinski02}).

A second possibility stems from the fact that the properties of the disk must
also change at this point.  For accretion to continue after zero-torque point,
the inner edge stops retreating, and the density at the inner edge must rise
\citep{Spruit93}.  In the scenario of FT07, this rise in inner disk density
would cause the magnitude of the soft lags to decrease again.  In the absence
of Compton upscattering, which would introduce hard lags not observed in
\saxj, their model predicts that the scattering from the disk will modify the
arrival times as $\Delta t(E) \approx (m_e c / \sigma_T n_e) (E_0^{-1} -
E^{-1})$, where $\sigma_T$ is the Thompson cross-section and $n_e$ is the
electron density of the scattering region.  This model provides reasonable
fits to the observed phase lags; Figure~\ref{fig:FalangaFitVsMDot} shows the
resulting $n_e$.

The fitted electron densities probe the relationship between the inner disk
and the accretion rate.  $n_e \sim 10^{19.5}$~cm$^{-3}$ gives a mean free path
of $\ell = (\sigma_T n_e)^{-1} \sim 0.5$~km and time $\ell/c \sim 2$~\us,
requiring $\sim$100 scattering events to produce the observed soft lags.  This
$\ell$ rules out the thin neutron star atmosphere as a scattering medium.  On
the other hand, simulations by \citet{Romanova04} suggest an inner disk
thickness of order the neutron star radius, and $(R / \ell)^2 \sim 400$ is
more than adequate to accomodate the number of needed scattering events.  From
the fits in Figure~\ref{fig:FalangaFitVsMDot}, we find that $n_e \propto
\dot{M}^{1.0\pm0.1}$ during the main outburst; during the presumed propeller
stage, $n_e \propto \dot{M}^{-0.4\pm0.1}$.  An interesting consequence of the
former relationship is that at higher accretion rates before the turnover,
when $r_m / r_c \lesssim 0.7$, the flow rate (the volume of material per unit
time crossing the inner edge of the disk) must be constant: since $\dot{M} =
\Omega m_p n_e r_m^2 v_r$, where $\Omega$ is the solid angle of the inner edge
and $v_r$ is the radial velocity of material crossing it, the flow rate
$\Omega r_m^2 v_r$ is fixed.

If the phase lags in \saxj\ are entirely due to Compton downscattering, as
suggested by FT07, then this model must explain why the magnitudes of the soft
lags for the second harmonic are roughly half the magnitudes for the
fundamental over a wide range of $\dot{M}$, as can be seen in
Figure~\ref{fig:PhaseLagSlopeVsFlux}.  Following the formalism of
\citet{Sunyaev80}, we model the timing effects of Comptonization with a
distribution $P(\Delta t)$ of the photon escape times from the scattering
medium.  For the $k$th harmonic of the spin frequency $\nu_s$, we convolve the
harmonic component $x_k(t) = A_k \exp(2\pi i k\nu_s t)$ with $P(\Delta t)$ to
find the new amplitude $A'_k$ and the time lag $\Delta t$ of the scattered
pulses:
\begin{equation}
  A_k'e^{2\pi i k\nu_s (t - \Delta t)} = 
    \int_{-\infty}^t A_k e^{2\pi i k\nu_s t'} P(t' - t) dt'
  \label{eq:ST80Formalism}
\end{equation}
In general, the choice of $P(\Delta t)$ accounts for the geometry of the
emitting and scattering regions.  However, for the large number ($\sim$100) of
scatterings previously noted, the geometry becomes unimportant and a photon
diffusion distribution of $P(\Delta t) = \tau^{-1} \exp(\Delta t/\tau)$ is
appropriate \citep{Sunyaev80}.  Applying this form to
equation~(\ref{eq:ST80Formalism}), the time lag of the $k^{\rm{th}}$ harmonic
is
\begin{equation}
  \Delta t_k = (2\pi k\nu_s\tau)^{-1} \tan^{-1} 2\pi k\nu_s\tau \, .
\end{equation}
The diffusion timescale cannot be longer than the lags themselves:  $\tau
\lesssim \Delta t_{\rm{max}} = 200$~\us.  This sets a lower limit of $\Delta
t_2 / \Delta t_1 \gtrsim 0.85$ due to scattering alone, contrary to the
observation that $\Delta t_2 / \Delta t_1 \approx 0.5$.  Thus Compton
scattering cannot by itself fully explain the observed lags: a change in the
energy dependence of the initial pulses is still necessary.

Clearly, detailed calculations are required to check whether the two models
really would predict the outcomes that we suggest, but they are at least in
principle plausible.  One would also need to test whether the models would
predict any reversal in other pulsation properties at propeller onset, given
that no such changes are observed.

The interesting behavior of the soft lags in this source suggests that the
lags of the other accretion-powered millisecond pulsars should be revisited to
check for flux dependence.  The only source for which this has been done is
XTE~J1814$-$338: an increase in soft lags was observed in the tail of the
outburst \citep{Watts06}, but the rapid drop off and shorter tail of this
source made it very difficult to study the pulsations at the lowest fluxes.
The other AMPs may be more promising as candidates for further study.

\acknowledgements

JMH thanks the Max Planck Institut f\"ur Astrophysik for hospitality that
facilitated the pursuit of this research.  We would also like to thank Lev
Titarchuk, Juri Poutanen, and J. Martin Laming for useful conversations.  This
work was supported at MIT in part by NASA grants under the RXTE Guest Observer
Program.


\begin{thebibliography}{38}
\expandafter\ifx\csname natexlab\endcsname\relax\def\natexlab#1{#1}\fi

\bibitem[{{Chakrabarty} \& {Morgan}(1998)}]{Chakrabarty98}
{Chakrabarty}, D. \& {Morgan}, E.~H. 1998, \nat, 394, 346

\bibitem[{{Cui} {et~al.}(1998){Cui}, {Morgan}, \& {Titarchuk}}]{Cui98}
{Cui}, W., {Morgan}, E.~H., \& {Titarchuk}, L.~G. 1998, \apjl, 504, L27

\bibitem[{{Di Salvo} {et~al.}(2008){Di Salvo}, {Burderi}, {Riggio}, {Papitto},
  \& {Menna}}]{DiSalvo08}
{Di Salvo}, T., {Burderi}, L., {Riggio}, A., {Papitto}, A., \& {Menna}, M.~T.
  2008, \mnras, 389, 1851

\bibitem[{{Falanga} {et~al.}(2005){Falanga}, {Kuiper}, {Poutanen}, {Bonning},
  {Hermsen}, {Di Salvo}, {Goldoni}, {Goldwurm}, {Shaw}, \&
  {Stella}}]{Falanga05}
{Falanga}, M., {Kuiper}, L., {Poutanen}, J., {Bonning}, E.~W., {Hermsen}, W.,
  {Di Salvo}, T., {Goldoni}, P., {Goldwurm}, A., {Shaw}, S.~E., \& {Stella}, L.
  2005, \aap, 444, 15

\bibitem[{{Falanga} \& {Titarchuk}(2007)}]{Falanga07}
{Falanga}, M. \& {Titarchuk}, L. 2007, \apj, 661, 1084 (FT07)

\bibitem[{{Ford}(2000)}]{Ford00}
{Ford}, E.~C. 2000, \apjl, 535, L119

\bibitem[{{Galloway} {et~al.}(2002){Galloway}, {Chakrabarty}, {Morgan}, \&
  {Remillard}}]{Galloway02}
{Galloway}, D.~K., {Chakrabarty}, D., {Morgan}, E.~H., \& {Remillard}, R.~A.
  2002, \apjl, 576, L137

\bibitem[{{Galloway} \& {Cumming}(2006)}]{Galloway06}
{Galloway}, D.~K. \& {Cumming}, A. 2006, \apj, 652, 559

\bibitem[{{Galloway} {et~al.}(2005){Galloway}, {Markwardt}, {Morgan},
  {Chakrabarty}, \& {Strohmayer}}]{Galloway05}
{Galloway}, D.~K., {Markwardt}, C.~B., {Morgan}, E.~H., {Chakrabarty}, D., \&
  {Strohmayer}, T.~E. 2005, \apjl, 622, L45

\bibitem[{{Galloway} {et~al.}(2007){Galloway}, {Morgan}, {Krauss}, {Kaaret}, \&
  {Chakrabarty}}]{Galloway07}
{Galloway}, D.~K., {Morgan}, E.~H., {Krauss}, M.~I., {Kaaret}, P., \&
  {Chakrabarty}, D. 2007, \apjl, 654, L73

\bibitem[{{Galloway} {et~al.}(2008){Galloway}, {Muno}, {Hartman}, {Savov},
  {Psaltis}, \& {Chakrabarty}}]{BurstDB}
{Galloway}, D.~K., {Muno}, M.~P., {Hartman}, J.~M., {Savov}, P., {Psaltis}, D.,
  \& {Chakrabarty}, D. 2008, \apjs, 179, 360

\bibitem[{{Ghosh} {et~al.}(1977){Ghosh}, {Pethick}, \& {Lamb}}]{Ghosh77}
{Ghosh}, P., {Pethick}, C.~J., \& {Lamb}, F.~K. 1977, \apj, 217, 578

\bibitem[{{Gierli{\'n}ski} {et~al.}(2002){Gierli{\'n}ski}, {Done}, \&
  {Barret}}]{Gierlinski02}
{Gierli{\'n}ski}, M., {Done}, C., \& {Barret}, D. 2002, \mnras, 331, 141

\bibitem[{{Gierli{\'n}ski} \& {Poutanen}(2005)}]{Gierlinski05}
{Gierli{\'n}ski}, M. \& {Poutanen}, J. 2005, \mnras, 359, 1261

\bibitem[{{Gilfanov} {et~al.}(1998){Gilfanov}, {Revnivtsev}, {Sunyaev}, \&
  {Churazov}}]{Gilfanov98}
{Gilfanov}, M., {Revnivtsev}, M., {Sunyaev}, R., \& {Churazov}, E. 1998, \aap,
  338, L83

\bibitem[{{Groth}(1975)}]{Groth75}
{Groth}, E.~J. 1975, \apjs, 29, 285

\bibitem[{{Hartman} {et~al.}(2008){Hartman}, {Patruno}, {Chakrabarty},
  {Kaplan}, {Markwardt}, {Morgan}, {Ray}, {van der Klis}, \&
  {Wijnands}}]{Hartman08}
{Hartman}, J.~M., {Patruno}, A., {Chakrabarty}, D., {Kaplan}, D.~L.,
  {Markwardt}, C.~B., {Morgan}, E.~H., {Ray}, P.~S., {van der Klis}, M., \&
  {Wijnands}, R. 2008, \apj, 675, 1468 (H08)

\bibitem[{{Ibragimov} \& {Poutanen}(2009)}]{Ibragimov09}
{Ibragimov}, A. \& {Poutanen}, J. 2009, \mnras, submitted, arXiv: 0901.0073

\bibitem[{{Illarionov} \& {Sunyaev}(1975)}]{Illarionov75}
{Illarionov}, A.~F. \& {Sunyaev}, R.~A. 1975, \aap, 39, 185

\bibitem[{{Jahoda} {et~al.}(2006){Jahoda}, {Markwardt}, {Radeva}, {Rots},
  {Stark}, {Swank}, {Strohmayer}, \& {Zhang}}]{Jahoda06}
{Jahoda}, K., {Markwardt}, C.~B., {Radeva}, Y., {Rots}, A.~H., {Stark}, M.~J.,
  {Swank}, J.~H., {Strohmayer}, T.~E., \& {Zhang}, W. 2006, \apjs, 163, 401

\bibitem[{{Jahoda} {et~al.}(1996){Jahoda}, {Swank}, {Giles}, {Stark},
  {Strohmayer}, {Zhang}, \& {Morgan}}]{Jahoda96}
{Jahoda}, K., {Swank}, J.~H., {Giles}, A.~B., {Stark}, M.~J., {Strohmayer}, T.,
  {Zhang}, W., \& {Morgan}, E.~H. 1996, in Proc. SPIE Vol. 2808, p. 59-70, EUV,
  X-Ray, and Gamma-Ray Instrumentation for Astronomy VII, Oswald H. Siegmund;
  Mark A. Gummin; Eds., ed. O.~H. {Siegmund} \& M.~A. {Gummin}, 59--70

\bibitem[{{Kirsch} {et~al.}(2004){Kirsch}, {Mukerjee}, {Breitfellner},
  {Djavidnia}, {Freyberg}, {Kendziorra}, \& {Smith}}]{Kirsch04}
{Kirsch}, M.~G.~F., {Mukerjee}, K., {Breitfellner}, M.~G., {Djavidnia}, S.,
  {Freyberg}, M.~J., {Kendziorra}, E., \& {Smith}, M.~J.~S. 2004, \aap, 423, L9

\bibitem[{{Long} {et~al.}(2005){Long}, {Romanova}, \& {Lovelace}}]{Long05}
{Long}, M., {Romanova}, M.~M., \& {Lovelace}, R.~V.~E. 2005, \apj, 634, 1214

\bibitem[{{Patruno} {et~al.}(2009){Patruno}, {Rea}, {Altamirano}, {Linares},
  {Wijnands}, \& {van der Klis}}]{Patruno09b}
{Patruno}, A., {Rea}, N., {Altamirano}, D., {Linares}, M., {Wijnands}, R., \&
  {van der Klis}, M. 2009, \mnras, accepted, arXiv: 0903.3210

\bibitem[{{Poutanen} \& {Gierli{\'n}ski}(2003)}]{Poutanen03}
{Poutanen}, J. \& {Gierli{\'n}ski}, M. 2003, \mnras, 343, 1301 (PG03)

\bibitem[{{Psaltis} \& {Chakrabarty}(1999)}]{Psaltis99}
{Psaltis}, D. \& {Chakrabarty}, D. 1999, \apj, 521, 332

\bibitem[{{Rappaport} {et~al.}(2004){Rappaport}, {Fregeau}, \&
  {Spruit}}]{Rappaport04}
{Rappaport}, S.~A., {Fregeau}, J.~M., \& {Spruit}, H. 2004, \apj, 606, 436

\bibitem[{{Romanova} {et~al.}(2004){Romanova}, {Ustyugova}, {Koldoba}, \&
  {Lovelace}}]{Romanova04}
{Romanova}, M.~M., {Ustyugova}, G.~V., {Koldoba}, A.~V., \& {Lovelace},
  R.~V.~E. 2004, \apj, 610, 920

\bibitem[{{Rubiano} {et~al.}(2004){Rubiano}, {Florido}, {Rodriguez}, {Gil},
  {Martel}, \& {Minguez}}]{Rubiano04}
{Rubiano}, J.~G., {Florido}, R., {Rodriguez}, R., {Gil}, J.~M., {Martel}, P.,
  \& {Minguez}, E. 2004, Journal of Quantitative Spectroscopy and Radiative
  Transfer, 83, 159

\bibitem[{{Sazonov} \& {Sunyaev}(2001)}]{Sazonov01}
{Sazonov}, S.~Y. \& {Sunyaev}, R.~A. 2001, \aap, 373, 241

\bibitem[{{Shakura} \& {Syunyaev}(1973)}]{Shakura73}
{Shakura}, N.~I. \& {Syunyaev}, R.~A. 1973, \aap, 24, 337

\bibitem[{{Spruit} \& {Taam}(1993)}]{Spruit93}
{Spruit}, H.~C. \& {Taam}, R.~E. 1993, \apj, 402, 593

\bibitem[{{Sunyaev} \& {Titarchuk}(1980)}]{Sunyaev80}
{Sunyaev}, R.~A. \& {Titarchuk}, L.~G. 1980, \aap, 86, 121

\bibitem[{{Vaughan} {et~al.}(1994){Vaughan}, {van der Klis}, {Wood}, {Norris},
  {Hertz}, {Michelson}, {van Paradijs}, {Lewin}, {Mitsuda}, \&
  {Penninx}}]{Vaughan94}
{Vaughan}, B.~A., {van der Klis}, M., {Wood}, K.~S., {Norris}, J.~P., {Hertz},
  P., {Michelson}, P.~F., {van Paradijs}, J., {Lewin}, W.~H.~G., {Mitsuda}, K.,
  \& {Penninx}, W. 1994, \apj, 435, 362

\bibitem[{{Watts} \& {Strohmayer}(2006)}]{Watts06}
{Watts}, A.~L. \& {Strohmayer}, T.~E. 2006, \mnras, 373, 769

\bibitem[{{Weinberg} {et~al.}(2001){Weinberg}, {Miller}, \&
  {Lamb}}]{Weinberg01}
{Weinberg}, N., {Miller}, M.~C., \& {Lamb}, D.~Q. 2001, \apj, 546, 1098

\bibitem[{{Wijnands} {et~al.}(2001){Wijnands}, {M{\'e}ndez}, {Markwardt}, {van
  der Klis}, {Chakrabarty}, \& {Morgan}}]{Wijnands01}
{Wijnands}, R., {M{\'e}ndez}, M., {Markwardt}, C., {van der Klis}, M.,
  {Chakrabarty}, D., \& {Morgan}, E. 2001, \apj, 560, 892

\bibitem[{{Wijnands} \& {van der Klis}(1998)}]{Wijnands98}
{Wijnands}, R. \& {van der Klis}, M. 1998, \nat, 394, 344

\end{thebibliography}
\end{document}